\documentclass[twoside]{article}
\usepackage{fleqn,espcrc2}
\usepackage{graphicx}

\newcommand{\AmS}{{\protect\the\textfont2
    A\kern-.1667em\lower.5ex\hbox{M}\kern-.125emS}}
\usepackage{here}
\def\beq{\begin{equation}}
\def\eeq{\end{equation}}
\def\bea{\begin{eqnarray}}
\def\eea{\end{eqnarray}}
\def\bq{\begin{quote}}
\def\eq{\end{quote}}

\def\nnb{\nonumber}
\def\ga{\left(}
\def\dr{\right)}
\def\aga{\left\{}
\def\adr{\right\}}

\def\rar{\rightarrow}
\def\lrar{\Longrightarrow}

\def\nnb{\nonumber}
\def\la{\langle}
\def\ra{\rangle}
\def\nin{\noindent}
\def\ba{\begin{array}}
\def\ea{\end{array}}

\title{\bf{\boldmath
$B^0_{d,s}$-$\bar B^0_{d,s}$ mass-differences
from QCD spectral sum rules}}
\author{Kaoru Hagiwara\address{KEK Theory Group, Tsukuba,
Ibaraki 305-0801, Japan
\\ E-mail: kaoru.hagiwara@kek.jp},
Stephan Narison\address{ Laboratoire
de Physique Math\'ematique, Universit\'e
de Montpellier II Place Eug\`ene
Bataillon, 34095 - Montpellier Cedex 05,
France \\ E-mail:
qcd@lpm.univ-montp2.fr}, and
Daisuke Nomura\address{Institute for Particle Physics Phenomenology,
University of Durham, Durham DH1 3LE, UK \\  E-mail:
daisuke.nomura@durham.ac.uk}}
\begin{document}
\pagestyle{empty}
\pagestyle{plain}
\begin{abstract}
\noindent
We present the {\it first} QCD spectral sum rules
analysis of the $SU(3)$ breaking parameter $\xi$ and an improved
estimate of the renormalization group invariant
(RGI) bag constant
$\hat{B}_{B_q}$ both entering into the
$B^0_{d,s}$-$\bar{B}^0_{d,s}$ mass-differences. The averages of
the results from the Laplace and moment sum rules to order $\alpha_s$
are
$f_B\sqrt{\hat B_B}\simeq (247 \pm 59)~{\rm MeV}$ and $
\xi\equiv {f_{B_s}\sqrt{\hat B_{B_s}}}/{f_{B}\sqrt{\hat B_{B}}}\simeq
(1.18\pm 0.03)$, in units where $f_\pi=130.7$ MeV.
  Combined with the experimental data on the mass-differences
$\Delta M_{d,s}$, one obtains the constraint on the CKM weak mixing angle
$|V_{ts}/V_{td}|^2\geq  20.0 (1.1)$. Alternatively, using the weak
mixing angle from the
analysis of the unitarity triangle and the data on $\Delta M_d$,
one predicts $\Delta M_s=18.6 (2.2)~ps^{-1}$ in agreement with
the present experimental lower bound and within
the reach of Tevatron 2.
\vspace*{1cm}
\\

\nin
May 2002\\
\end{abstract}
\maketitle
\section{Introduction}
\nin
    $ B^0_{(s)}$ and $\bar{B}^0_{(s)}$ are not eigenstates of the weak
Hamiltonian, such that their
oscillation frequency is governed by their mass-difference $\Delta
M_q$. The measurement
by the UA1 collaboration~\cite{UA1} of a large value of $\Delta M_d$
was the {\it first} indication of the heavy top quark mass.
In the SM, the mass-difference
is approximately given by~\cite{BURAS}
\begin{eqnarray}
\label{deltam}
\Delta M_q&\simeq& \frac{G^2_F}{4\pi^2}M^2_W|V_{tq}V^*_{tb}|^2
S_0\ga\frac{m^2_t}{M^2_W}\dr\eta_BC_B(\nu)\nnb\\
&&\times\frac{1}{2M_{B_q}}
\la \bar{B}^0_q|{\cal O}_q (x)|B^0_q\ra~ ,
\end{eqnarray}
where the $\Delta B=2$ local operator ${\cal O}_q(x)$ is defined as
\begin{eqnarray}
\label{eq:operator}
{\cal O}_q(x)\equiv (\bar b\gamma_\mu L q)(\bar b\gamma_\mu L q)~,
\end{eqnarray}
with  $L\equiv (1-\gamma_5)/2$ and $q \equiv d, s$.
$S_0,~\eta_B$ and $C_B(\nu)$ are short distance quantities calculable 
perturbatively.
(Here we are following the notation of Ref.~\cite{BURAS}.)
On the other hand, the matrix element $\la
\bar{B}^0_q|{\cal O}_q|B^0_q\ra$ requires
non-perturbative QCD calculations, and is usually parametrized for
as
\begin{eqnarray}
\la \bar{B}^0_q|{\cal O}_q|B^0_q\ra={4\over 3}
\frac{f^2_{B_q}}{2} M^2_{B_q}B_{B_q}~.
\end{eqnarray}
$f_{B_q} $ is the $B_q$ decay constant normalized as $f_\pi=130.7$ MeV, and
$B_{B_q}$ is the so-called bag parameter which is $B_{B_q} = 1$
if one uses a vacuum saturation of the matrix element and equal to 3/4 in the
large $N_c$ limit.
   From Eq.~(\ref{deltam}), it is clear that the measurement of $\Delta
M_d$ provides one of the CKM mixing angles $|V_{td}|$ if one uses
$|V_{tb}|\simeq 1$. One can
also extract this quantity from the ratio
\begin{eqnarray} \label{mass}
\frac{\Delta M_s}{\Delta M_d}
&=&
\left| \frac{V_{ts}}{V_{td}} \right|^2
\frac{M_{B_d}}{M_{B_s}}
\frac{\la \bar{B}^0_s|{\cal O}_s|B^0_s\ra}
       {\la \bar{B}^0_d|{\cal O}_d|B^0_d\ra}\nnb\\
&\equiv&
\left| \frac{V_{ts}}{V_{td}} \right|^2
   \frac{M_{B_d}}{M_{B_s}}\xi^2~,
\end{eqnarray}
since in the SM with three generations and unitarity constraints,
$|V_{ts}|\simeq |V_{cb}|$. Here
\begin{eqnarray} \label{bbs}
\xi
\equiv
\sqrt{\frac{g_s}{g_d}}
\equiv
\frac{f_{B_s}\sqrt{B_{B_s}}}{f_{B}\sqrt{B_{B}}}~.
\end{eqnarray}
The
great advantage of Eq.~(\ref{mass}) compared with the former relation
in Eq.~(\ref{deltam}) is
that in the ratio, different systematics in the evaluation of the
matrix element tends to cancel
out, thus providing a more accurate prediction. However, unlike
$\Delta M_d= 0.479(12)~ps^{-1}$, which is measured
with a good precision~\cite{PDG}, the
determination of
$\Delta M_s$ is an experimental challenge due to the rapid
oscillation of the $B^0_s$-$\bar{B}^0_s$
system. At present, only a lower bound of 13.1 $ps^{-1}$ is
available at the 95\% CL from experiments~\cite{PDG},
but this bound already provides a strong constraint on $|V_{td}|$.
\section{Two-point function sum rule}
\nin
Ref.~\cite{PICH} has extended the analysis of the
$K^0$-$\bar{K}^0$ systems of Ref.~\cite{RAFAEL}, using two-point
correlator of the four-quark operators into
the analysis of the quantity
$f_{B}\sqrt{B_{B}}$ which governs the
$B^0$-$\bar{B}^0$ mass difference. The two-point correlator defined as
\begin{eqnarray} \label{twopoint}
\psi_H(q^2) \equiv i \int d^4x ~e^{iqx} \
\la 0 | {\cal T}
{\cal O}_q(x) \ga {\cal O}_q(0)\dr ^\dagger | 0 \ra ~,
\end{eqnarray}
is built from the $\Delta B=2$ weak operator given
in Eq.~(\ref{eq:operator}).
The two-point function approach is very convenient due to
its simple analytic properties which are not the case of approach
based on three-point functions~\footnote{
For detailed criticisms, see \cite{SNB}.}. However, it
involves non-trivial QCD
calculations which become technically complicated when one includes
the contributions of radiative
corrections due to non-factorizable diagrams. These perturbative (PT)
radiative corrections
due to {\it factorizable and non-factorizable} diagrams  have been already
computed in Ref.~\cite{PIVO} (referred as NP), where it has been found
that the factorizable corrections
are large while the non-factorizable ones are negligibly small. NP
analysis has confirmed the
estimate in Ref.~\cite{PICH} from lowest order calculations, where under
some assumptions on the
contributions of higher mass resonances to the spectral function, the
value of the bag parameter $B_B$ has been found to be
\begin{eqnarray}\label{eq:pivo1}
B_{B_d}(4M_b^2)\simeq (1\pm 0.15)~.
\end{eqnarray}
This value is comparable with the one $B_{B_d}= 1$ from the
vacuum saturation estimate, which is expected to be a quite good
approximation due to the relative high-scale of the $B$-meson mass.
Equivalently, the corresponding RGI quantity is
\begin{eqnarray}
\label{eq:pivo2}
\hat B_{B_d}\simeq (1.5\pm 0.2),
\end{eqnarray}
where we have used the relation
\begin{eqnarray}
{\hat B_{B_q}}
=
    B_{B_q}(\nu)
{\alpha_s^{\frac{\gamma_0}{\beta_1}}}
\aga
       1 + \ga \frac{5165}{12696} \dr \ga \frac{\alpha_s}{\pi} \dr
\adr,
\end{eqnarray}
with $\gamma_0=1$ being the anomalous dimension of the
operator ${\cal O}_q$ and $\beta_1=-23/6$ for 5 flavours.
$\nu$ is the subtraction point.
The NLO corrections have been obtained in the
$\overline{MS}$ scheme~\cite{BURAS}.
We have also used the value of the bottom quark
pole mass~\cite{SNMAS,SNB}
\begin{eqnarray}
   M_b = (4.66\pm 0.06)~{\rm GeV}~.
\end{eqnarray}
In the following, we study {\it (for the first time)}, from the QCD spectral
sum rules (QSSR) method \footnote{Our preliminary results have been
presented in \cite{HEPMAD}.}, the
$SU(3)$ breaking effects on the ratio
$\xi$ defined previously in Eq.~(\ref{bbs}),
where a similar analysis of the ratios of the decay constants has
given the values~\cite{SNFBS}
\begin{eqnarray}
\label{fbs}
\frac{f_{D_s}}{f_{D}}\simeq 1.15\pm 0.04~,~~~
\frac{f_{B_s}}{f_{B}}\simeq 1.16\pm 0.04 .
   \end{eqnarray}
We shall also improve the previous result of Ref.~\cite{PICH,PIVO} on
$B_{B_d}$ by the inclusion of the
$B_q B_q^{*}$ and $B_q^{*} B_q^{*}$ resonances into the spectral
function.

\section{Inputs for the sum rule analysis}
\nin
We shall be concerned here with the two-point correlator defined
in Eq.~(\ref{twopoint}).
The hadronic part of the spectral function can be
conveniently parametrized using the effective realization \cite{PICH}
\begin{eqnarray}
{\cal O}^{\mbox{\scriptsize \it eff}}_q=\frac{1}{3} g_q
\partial_\mu B^0_q \partial^\mu B^0_q +...~,
\end{eqnarray}
where $...$ indicates higher resonances
and $ g_q \equiv f^2_{B_q} B_{B_q}$.
Retaining the $B B$, $B B^*$ and $B^* B^*$ resonances and
parametrizing the higher resonances with the QCD continuum
contribution, it gives
\begin{eqnarray}
&&\frac{1}{\pi}{\rm Im}\psi^{\mbox{\scriptsize \it eff}}_q(t) =
%
%
    \frac{2}{9} \left( \frac{g_q}{8\pi} \right)^2 t^2\times\Bigg{[} 
\nnb\\
&& \left( 1 - \frac{2 M^2_{B_q}}{t} \right)^2
     \sqrt{ 1 - \frac{4M^2_{B_q}}{t} }
     \theta( t - 4 M^2_{B_q} )                                      \nnb\\
%
%
&&
+
\left( 1 -  4 \frac{M^2_{B^\ast_q}}{t}
              + 12 \frac{M^4_{B^\ast_q}}{t^2} \right)
\times          \nnb\\
&&
     \sqrt{ 1 - \frac{4M^2_{B^\ast_q}}{t} }
     \theta( t - 4 M^2_{B^\ast_q} )                   \nnb\\
%
%
&&
+
2\lambda^{3/2}
     \left( 1 , \frac{M^2_{B_q}}{t}, \frac{M^2_{B^\ast_q}}{t} \right)
\times          \nnb\\
&&
     \theta ( t - ( M_{B_q} + M_{B^\ast_q} )^2 )\Bigg{]}
\end{eqnarray}
below the QCD continuum threshold $t_c$.
The function $\lambda(x,y,z)$ is a phase space factor,
\begin{eqnarray}
     \lambda(x,y,z) \equiv x^2 + y^2 + z^2 - 2 xy - 2 yz - 2 zx .
\end{eqnarray}
We have used, to a first approximation, the large $M_b$ and vacuum 
saturation relations:
\beq
g_q\equiv f^2_{B_q} B_{B_q}\simeq g_{q^*}\equiv  f^2_{B^*_q} B_{B^*_q}
\eeq
among the couplings. The results $f_B\approx f_{B^*}$ have been also obtained
from QCD spectral sum rules \cite{SNB}, while the vacuum saturation 
relation $B_{B^*_q}\simeq 1\simeq B_{B_q}$
is {\it a posteriori} expected to be a good approximation as 
indicated by the result obtained later on in
this paper.
The short distance expression of the spectral function is
obtained using the Operator Product Expansion (OPE) including
non-perturbative condensates~\cite{SVZ}~\footnote{
We shall not include here the effects
of tachyonic gluon mass~\cite{ZAK} or some other renormalon-like
terms which give small effects
in various examples~\cite{CNZ,SNI}.}.
The massless ($m_q=0$) expression for the lowest perturbative
and gluon condensate contributions has been obtained in
Ref.~\cite{PICH}.  Radiative
{\it factorizable and non-factorizable} corrections to the perturbative
graphs in the massless light
quark case have been obtained in NP~\cite{PIVO}.

\section{$SU(3)$ breaking contributions}
\nin
The lowest order perturbative contribution for
$m_s \neq 0$ to the two-point correlator is
\begin{eqnarray}
&& \frac{1}{\pi}{\rm Im} \psi^{pert}_s(t)
= \theta(t-4(M_b+m_s)^2)  \nonumber\\
&&
\times \frac{t^4}{1536 \pi^6}      \nonumber\\
&&
\times
\int_{(\sqrt{\delta}+\sqrt{\delta^\prime})^2}
       ^{(1-\sqrt{\delta}-\sqrt{\delta^\prime})^2} dz
\int_{(\sqrt{\delta}+\sqrt{\delta^\prime})^2}
^{(1-\sqrt{z})^2}      du  ~ zu
\nonumber \\
&&
\times
\lambda^{1/2} (1,z,u)~
\lambda^{1/2} \left(1,  \frac{\delta}{z},  \frac{\delta^\prime}{z} \right)
\lambda^{1/2} \left(1,  \frac{\delta}{u},  \frac{\delta^\prime}{u} \right)
\nonumber \\
&&
\times
\left[
    4 f \left(\frac{\delta}{z},  \frac{\delta^\prime}{z} \right)
      f \left(\frac{\delta}{u},  \frac{\delta^\prime}{u} \right)
\right. \nonumber\\
&&
\phantom{\times}
-2 f \left(\frac{\delta}{z},  \frac{\delta^\prime}{z} \right)
      g \left(\frac{\delta}{u},  \frac{\delta^\prime}{u} \right)
\nonumber\\
&& \phantom{\times}
-2 g \left(\frac{\delta}{z},  \frac{\delta^\prime}{z} \right)
      f \left(\frac{\delta}{u},  \frac{\delta^\prime}{u} \right)
    \nonumber\\
&& \phantom{\times} \left.
+
\frac{(1-z-u)^2}{zu}
      g \left(\frac{\delta}{z},  \frac{\delta^\prime}{z} \right)
      g \left(\frac{\delta}{u},  \frac{\delta^\prime}{u} \right)
\right] .
\end{eqnarray}
Here $\delta \equiv M_b^2/t$ and $\delta^\prime \equiv m_s^2/t$,
respectively.
The functions $f(x,y)$ and $g(x,y)$ are defined by
\begin{eqnarray}
&& f(x,y) \equiv 2 - x - y - (x-y)^2 ,\\
&& g(x,y) \equiv 1 + x + y -2(x-y)^2 .
\end{eqnarray}
We include the ${\cal O}(\alpha_s)$ correction
from factorizable diagrams by using
the results in the $\overline{MS}$ scheme for the two-point correlators of
currents~\cite{DJOUADI} \footnote{We shall neglect the 
nonfactorizable corrections in our analysis, according to
the results in NP \cite{PIVO} obtained in a slight variant of the 
$\overline{MS}$ scheme.}
This can be done using the convolution formula,
\begin{eqnarray}
&&
    \frac1{\pi} {\rm Im} \psi^{\alpha_s}_{s} (t)
= \theta(t-4(M_b+m_s)^2)
\nonumber \\
&&
\times \frac{t^2}{6\pi^4}
\int_{(\sqrt{\delta}+\sqrt{\delta^\prime})^2}
    ^{(1-\sqrt{\delta}-\sqrt{\delta^\prime})^2} dz
\int_{(\sqrt{\delta}+\sqrt{\delta^\prime})^2}^{(1-\sqrt{z})^2} du
\nonumber \\
&&
\times \lambda^{1/2} (1,z,u)
\nonumber \\
&&
\times
\left\{
     {\rm Im} \Pi^0_{\mu\nu}(zt) {\rm Im} \Pi^{\alpha_s \mu\nu}(ut)
\right.
\nonumber \\
&&
\left.
\phantom{\times}
+ {\rm Im} \Pi^{\alpha_s}_{\mu\nu}(zt) {\rm Im} \Pi^{0 \mu\nu}(ut)
\right\} .
\end{eqnarray}
Here $\Pi^0_{\mu\nu}(q^2)$ and $\Pi^{\alpha_s}_{\mu\nu}(q^2)$ are
respectively the lowest and the next-to-leading order QCD contribution
to the two point correlator $\Pi_{\mu\nu}(q^2)$ defined by
\begin{eqnarray}
&& \Pi_{\mu\nu}(q^2) \equiv
i \int d^4 x ~e^{iqx} \nonumber\\
&&
    \times
    \langle 0|T ( \bar{b} (x) \gamma_\mu L s (x) )
                ( \bar{s} (0) \gamma_\nu L b (0) ) |0 \rangle .
\end{eqnarray}
The quark condensate contribution reads
\begin{eqnarray}
&& \frac{1}{\pi}{\rm Im}\psi^{\langle \bar{s} s \rangle}_s(t) =
\nonumber \\
&&
\theta( t - 4 ( M_b + m_s )^2 ) \frac{1}{384\pi^3}
m_s \langle \bar{s} s \rangle
\nonumber \\
&&
\times
\int_{( M_b + m_s )^2}^{ \left( \sqrt{t} - M_b \right)^2}  d q_1^2~
\sqrt{\lambda_1}
\left(
    4 + 2 q^2 \frac{\partial}{\partial q^2}
\right)
\nonumber \\
&&
\left[
\sqrt{\lambda_0}
\left\{
     \lambda_1 \left( 1 + \frac{M_b^2}{q^2} - \frac{q_1^2}{q^2} \right)
      q_1^2
\right. \right.
\nonumber \\
&&
\left. \left.
+    f_1    \left( 1 - \frac{M_b^2}{q^2} + \frac{q_1^2}{q^2} \right)
               \left( q^2 - M_b^2 - q_1^2 \right)
\right\}
\right] .
\end{eqnarray}
Here $\lambda_0$, $\lambda_1$, and $f_1$ are defined by
\begin{eqnarray}
    \lambda_0 &\equiv&
       \lambda \left( 1, \frac{q_1^2}{q^2}, \frac{M_b^2}{q^2} \right),  \\
    \lambda_1 &\equiv&
       \lambda \left( 1, \frac{M_b^2}{q_1^2}, \frac{m_s^2}{q_1^2} \right), \\
    f_1 &\equiv&
         1 + \frac{M_b^2}{q_1^2} + \frac{m_s^2}{q_1^2}
       - 2  \frac{( M_b^2 - m_s^2 )^2}{q_1^4}  .
\end{eqnarray}
\begin{figure}[hbt]
\begin{center}
\includegraphics[width=8cm]{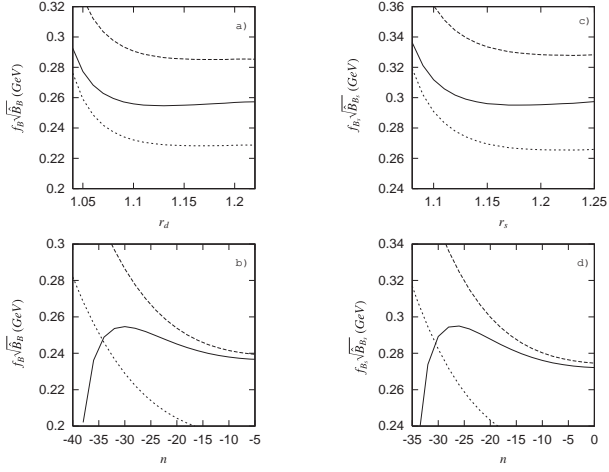}
\caption{\footnotesize{Moment sum rules analysis of
$f_{B_{(s)}}\sqrt{\hat{B}_{B_{(s)}}}$ for different
values of $r_q$ and $n$:
$f_{B}\sqrt{\hat{B}_{B}}$ versus: a) $r_d$ at $n=-30$,
b) $n$ at $r_d=1.13$; $f_{B_s}\sqrt{\hat{B}_{B_s}}$ versus: c) $r_s$ 
at $n=-26$,
d) versus $n$ at $r_s=1.17$.
Dotted curve: lowest order
perturbative contribution; dashed curve:
lowest order perturbative + $m_s\langle \bar{s}s \rangle$
[only for c) and d)] + $\la \alpha_s G^2\ra$ condensates.
Solid curves: total contribution to order $\alpha_s$.}}
\label{moments}
\end{center}
\end{figure}
\section{The sum rule analysis}
\begin{figure}[hbt]
\begin{center}
\includegraphics[width=8cm]{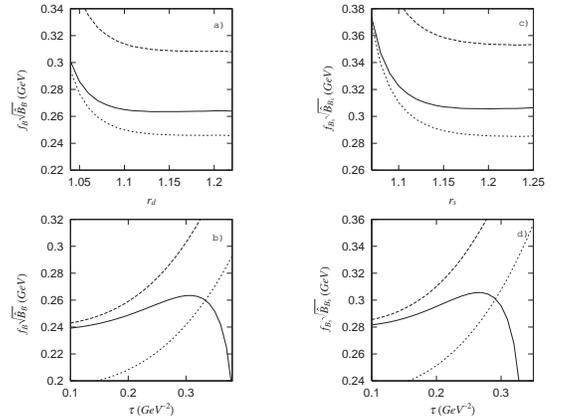}
\caption{\footnotesize{Laplace sum rules analysis of
$f_{B_{(s)}}\sqrt{\hat{B}_{B_{(s)}}}$ for different values of $r_q$ and $\tau$:
$f_{B}\sqrt{\hat{B}_{B}}$ versus: a) $r_d$ at  $\tau=0.31 {\rm GeV}^{-2}$,
b) $\tau$
at $r_d=1.13$;  $f_{B_s}\sqrt{\hat{B}_{B_s}}$ versus: c) $r_s$
at $\tau=0.26 {\rm GeV}^{-2}$,
d) versus $\tau$
at $r_s=1.20$.
The curves are the same as
in Fig.~\ref{moments}.}}
\label{laplace}
\end{center}
\end{figure}

\nin
For the sum rule analysis, we shall work
like \cite{PICH,PIVO} with the moments
\begin{eqnarray}
{\cal M}^{(n)}_q=
    \int_{4(M_b+m_q)^2}^{t_c^{(q)}} dt~t^n~\frac{1}{\pi}
    {\rm Im} \psi_q(t)~,
\end{eqnarray}
and with the Laplace sum rule
\begin{eqnarray}
{\cal L}{(\tau)}_q=
    \int_{4(M_b+m_q)^2}^{t_c^{(q)}}dt~e^{-t\tau}~\frac{1}{\pi}
    {\rm Im}\psi_q(t)~.
\end{eqnarray}
In so doing, in addition to the pQCD input parameters given previously, we
shall need the values of the QCD condensates and $SU(3)$ breaking
parameters, which we give in Table 1. We show in Fig.~\ref{moments} 
the moment sum rules
analysis of $f_{B_{(s)}}\sqrt{\hat{B}_{B_{(s)}}}$ for different
values of $r_q$ and $n$. As one can see from Fig.~\ref{moments}a, b, 
the stability regions of
the quantity,
\begin{eqnarray}
   \sqrt{\hat g_d}\equiv f_{B_d}\sqrt{\hat B_{B_d}}
\end{eqnarray}
versus the number $n$ of moments and the continuum threshold
\begin{eqnarray}
    {r_d}\equiv \frac{t^{(d)}_c}{4 M_b^2}~,
\end{eqnarray}
are obtained for large ranges ending with an extremum for
\begin{eqnarray}
    n\simeq -30~,~~~~~ r_d\simeq 1.13~.
\end{eqnarray}
These range of values of the sum rule parameters are in good agreement
with previous results in Ref.~\cite{PICH} and NP \cite{PIVO}. Analogous
values of $n$ and $r_s$ stabilities
are also obtained in the analysis of
$\sqrt{\hat{g}_s} \equiv f_{B_s} \sqrt{\hat{B}_{B_s}}$
(see Fig.~\ref{moments}c, d), with
\begin{eqnarray}
    n\simeq -26~,~~~~~ r_s\simeq 1.17~.
\end{eqnarray}
One can notice that the stabilities in the continuum for $\hat{g}_d$
and $\hat{g}_s$
differ slightly as a reflection of the $SU(3)$ breakings, which one
can parametrize numerically as
\begin{eqnarray}
    r_s\simeq \ga\sqrt{r_d}+\frac{m_s}{M_b}\dr^2 \approx r_d+0.05~.
\end{eqnarray}
Similar analysis is done with the
Laplace sum rules. We show in Fig.~\ref{laplace} the predictions of
$f_{B_{(s)}}\sqrt{\hat{B}_{B_{(s)}}}$ for different values of $r_q$ 
and $\tau$, where an extremum is obtained
for
\beq
\tau \simeq 0.3~{\rm GeV}^{-2}~.
\eeq

%
\section{Results and implications on \boldmath $|V_{ts}/V_{td}|^2$ and
\boldmath$\Delta M_s$}
\begin{table*}[hbt]
\setlength{\tabcolsep}{.pc}
\newlength{\digitwidth} \settowidth{\digitwidth}{\rm 0}
\catcode`?=\active \def?{\kern\digitwidth}
\caption{\footnotesize{Different sources of errors in the estimate of
$f_B\sqrt{\hat{B}_B}$ and
$\xi$ in units where $f_\pi=130.7$ MeV. The box marked with -- means
that the error is zero or negligible.}}
\begin{center}
\begin{tabular*}{\textwidth}{@{}l@{\extracolsep{\fill}}rrrr}
\hline \\
Sources&\multicolumn{2}{c}{$\Delta \ga f_B\sqrt{\hat{B}_B}\dr$
[MeV]}&\multicolumn{2}{c}{$\Delta\xi\times 10^2$}\\
\\
\hline\\
&Moments&Laplace&Moments&Laplace\\
\cline{2-3}\cline{4-5}
&\\
$-n\simeq (30\sim 10)$                      &  8.3 &  --  & 1.5 & --  \\
$\tau\simeq (0.1\sim 0.31)~{\rm GeV}^{-2}$  & --   & 12.4 &  -- & 1.7 \\
$r_d\simeq 1.06\sim 1.17$                   &  7.9 &  7.8 & 1.0 & 2.5 \\
$\Lambda_5=(216 ^{+25}_{-24})$~MeV
              \cite{PDG,BETHKE}              & 0.4  &  0.4 & 0.1 & 0.1 \\
$\nu=M_b\sim 2M_b$                          &  8.7 &  9.1 & 0.2 & 0.2 \\
$\alpha_s^2$: geometric PT series           & 43.0 & 45.0 & 0.6 & 0.7 \\
$M_b=(4.66\pm 0.06)$~{GeV} \cite{SNMAS}     & 34.6 & 37.1 & 0.9 & 1.2 \\
$\la\alpha_s G^2\ra=(0.07\pm 0.01)~{\rm GeV}^4$
                                     \cite{SNG}& 1.3 & 1.6 & --  & -- \\
$\la\bar uu\ra(2)=-(254\pm 15)^3$ MeV$^3$
                            \cite{SNMAS,SNL,SNB} & -- & -- &  0.2 & 0.3 \\
$\la\bar ss\ra/\la\bar uu\ra=0.7\pm 0.2$ \cite{SNMAS,SNL,SNB}&--&--&0.3&0.4\\
$\overline{m}_s(2)=(117\pm 23)$ MeV \cite{SNMAS,SNL,SNB} & -- & -- & 
1.5 & 1.7 \\
\\
\hline \\
Total & 57.1 & 60.8 & 2.6 & 3.8 \\ \\
\hline
\end{tabular*}
\end{center}
\end{table*}
\nin
We take as a conservative result for $\hat{g}_d$, from the moments 
sum rule analysis, the one from a large range
of $n=-10$ to $-30$ and for
$r_d=1.06$ to 1.17. Adding quadratically the different sources of
errors in Table 1, we obtain
\begin{eqnarray}\label{fbb}
f_B\sqrt{\hat B_B}\simeq (245 \pm 57)~{\rm MeV}~,
\end{eqnarray}
in units where $f_\pi=130.7$ MeV.
The most relevant errors given in Eq.~(\ref{fbb})
come from $M_b$ and the truncation of the PT series.
We have estimated the latter by assuming that the coefficient of the
$\alpha_s^2$ contribution
comes from a geometric growth of the PT coefficients. The other
parameters $n, r_d, \Lambda,~\la\bar qq\ra, ~\la \alpha_s
G^2\ra$ and $\nu$ (subtraction point) induce smaller errors as given
in Table 1. We proceed in a similar way for
$\hat{g}_s$. Then, we take the range $n=-10$ to $-26$ and $r_s= 1.10$ to
1.21, and deduce the ratio
\begin{eqnarray}
\xi \equiv\frac{f_{B_s}\sqrt{B_{B_s}}}{f_{B}\sqrt{B_{B}}}\simeq
1.174\pm 0.026~,
\end{eqnarray}
where the errors come almost equally from
$n,~r_q,~m_s,~M_b$ and the $\alpha_s^2$ term.
As expected, we have smaller errors for the ratio $\xi$ due to the
cancellation of the systematics. \\ We proceed in the same way with the
Laplace sum rules where we take the range of $\tau$ values from 0.1
to 0.37 GeV$^{-2}$ (see Fig.~\ref{laplace} a, b) in order to have a 
conservative
result. Then, we
deduce
\begin{eqnarray}
\label{xi-laplace}
f_B\sqrt{\hat B_B}&\simeq& (249 \pm 61)~{\rm MeV}~, \nnb\\
\xi &\simeq& 1.187\pm 0.038~.
\end{eqnarray}
As a final result, we take the arithmetic average from the moments and
Laplace sum rules results. Then, we deduce
\begin{eqnarray}
\label{xi-averaged}
   f_B\sqrt{\hat B_B} &\simeq& (247 \pm 59)~{\rm MeV},\nnb\\
                  \xi &\simeq& 1.18\pm 0.03~,
\end{eqnarray}
in the unit where $f_\pi=130.7$ MeV.
These results can be compared  with different lattice $f_B\sqrt{\hat 
B_B} \simeq (230 \pm
32)$~{\rm MeV}, $\xi\simeq 1.14\pm 0.06$, and global-fit of the CKM 
mixing angles giving $f_B\sqrt{\hat B_B}
\simeq (231\pm 15)$ MeV quoted in \cite{LATT,LATT2}. By comparing our 
results Eq.~(\ref{xi-averaged})
with the one of $f_{B_s}/f_{B_d}$ in Eq.~(\ref{fbs})
\footnote{One can notice that similar strengths of the $SU(3)$
breakings have been obtained
for the $B\rar K^*\gamma$ and $B\rar K l\nu$ form factors
\cite{SNFORM}.}, one can conclude
(to a good approximation) that
\begin{eqnarray}
\hat{B}_{B_s}\approx \hat{B}_{B_d}
   &\simeq&
  (1.65\pm 0.38)~~~\lrar\nnb\\
  B_{B_{d,s}}(4 M^2_b)
&\simeq&
(1.1\pm 0.25)~,
\end{eqnarray}
indicating a negligible $SU(3)$ breaking for the bag parameter. For a
consistency, we have used the
estimate to order $\alpha_s$~\cite{SNFB1}
\begin{eqnarray}
f_B \simeq (1.47\pm 0.10)f_\pi\simeq (192\pm 19)~{\rm MeV}~,
\end{eqnarray}
and we have assumed that the error from $f_B$ compensates the one in
Eq.~(\ref{xi-averaged}). The result is
in excellent agreement with the previous result
of Ref.~\cite{PIVO} in Eqs.~(\ref{eq:pivo1}) and (\ref{eq:pivo2}), and
agrees within the errors with the lattice estimates \cite{LATT,LATT2}.\\
    Using the experimental values
\begin{eqnarray}
\Delta M_d&=& 0.479(12)~ps^{-1}~,\nnb\\
\Delta M_s&\geq& 13.1~ps^{-1}~~(95\%~{\rm CL}),
\end{eqnarray}
one can deduce from Eq.~(\ref{mass})
\begin{eqnarray}\label{rhosd}
   \rho_{sd}
\equiv
   \left| \frac{V_{ts}}{V_{td}} \right|^2 \geq 20.0 (1.1).
\end{eqnarray}
Alternatively, using
\begin{eqnarray}
\rho_{sd}\simeq
\frac{1}{\lambda^2\big{[}(1-\bar\rho)^2+\bar\eta^2\big{]}}\simeq
28.4(2.9)
\end{eqnarray}
with \cite{LATT}
\begin{eqnarray}
\lambda&\simeq&0.2237(33)~,~~~\nnb\\
\bar\rho&\equiv&\rho\ga 1-\frac{\lambda^2}{2}\dr\simeq 0.223(38)~,\nnb\\
\bar\eta&\equiv&\eta\ga 1-\frac{\lambda^2}{2}\dr\simeq 0.316(40)~,
\end{eqnarray}
$\lambda,~\rho$ and $\eta$ being the Wolfenstein parameters,
    we deduce
\begin{eqnarray}\label{massdiff}
    \Delta M_s\simeq 18.6(2.2)~ps^{-1}~,
\end{eqnarray}
in agreement with the present experimental lower bound and
within the reach of Tevatron run~2
experiments.
\section{Conclusions}
\nin
We have applied QCD spectral sum rules for extracting {\it (for the
first time)}
the $SU(3)$ breaking parameter $\xi$, and for improving the
estimate of the quantity $f_B\sqrt{\hat B_B}$. Our predictions are 
given in Eq.~(\ref{xi-averaged}).
The phenomenological consequences of our results for the
CKM mixing angle and $B^0_{d,s}$-$\bar B^0_{d,s}$ mass-differences
are given in Eqs. (\ref{rhosd}) and (\ref{massdiff}) .

\section*{Acknowledgements}
\nin
This work has been initiated when one of the authors (SN) has been a visiting
professor at the KEK theory group from January to July 2000.

\end{document}